\documentclass[12pt]{article}
\usepackage{amssymb}
\begin{document} 

\def\squeeze{\hskip-17pt}
\def\kbar{{\mathchar'26\mkern-9muk}}  
\def\bra#1{\langle #1 \vert}
\def\ket#1{\vert #1 \rangle}
\def\vev#1{\langle #1 \rangle}
\def\tr{\mbox{Tr}\,}
\def\ad{\mbox{ad}\,}
\def\ker{\mbox{Ker}\,}
\def\im{\mbox{Im}\,}
\def\ad{\mbox{ad}\,}
\def\Dirac{{\raise0.09em\hbox{/}}\kern-0.69em D}
\def\b#1{{\mathbb #1}}
\def\c#1{{\cal #1}}
\def\etal{{\it et al.}\ }
\def\st{\;\vert\;}
\def\der{\mbox{Der}}

\newcommand{\sect}[1]{\setcounter{equation}{0}\section{#1}}
\renewcommand{\theequation}{\thesubsection.\arabic{equation}}
\renewcommand{\theequation}{\thesection.\arabic{equation}}

\newcommand{\be}{\begin{equation}}
\newcommand{\ee}{\end{equation}}
\newcommand{\bea}{\begin{eqnarray}}
\newcommand{\eea}{\end{eqnarray}}

\title{Noncommutative Geometry for Pedestrians\footnote{Lecure given
       at the International School of Gravitation, Erice: 16th Course: 
       `Classical and Quantum  Non-Locality'.}}

\author{J. Madore \\Laboratoire de Physique Th\'eorique\\
        Universit\'e de Paris-Sud, B\^atiment 211, F-91405 Orsay
   \and Max-Planck-Institut f\"ur Physik\\
        F\"ohringer Ring 6, D-80805 M\"unchen
       }

\date{}

\maketitle

\abstract{A short historical review is made of some recent literature in
the field of noncommutative geometry, especially the efforts to add a
gravitational field to noncommutative models of space-time and to use it
as an ultraviolet regulator. An extensive bibliography has been added
containing reference to recent review articles as well as to part of the
original literature.}

\parskip 4pt plus2pt minus2pt
\vfill
\noindent
LMU-TPW 99-11
\newpage

\sect{Introduction}

To control the divergences which from the very beginning had plagued
quantum electrodynamics, Heisenberg already in the 1930's proposed to
replace the space-time continuum by a lattice structure.  A lattice
however breaks Lorentz invariance and can hardly be considered as
fundamental. It was Snyder~\cite{Sny47a,Sny47b} who first had the idea
of using a noncommutative structure at small length scales to introduce
an effective cut-off in field theory similar to a lattice but at the
same time maintaining Lorentz invariance.  His suggestion came however
just at the time when the renormalization program finally successfully
became an effective if rather {\it ad hoc} prescription for predicting
numbers from the theory of quantum electrodynamics and it was for the
most part ignored. Some time later von~Neumann introduced the term
`noncommutative geometry' to refer in general to a geometry in which an
algebra of functions is replaced by a noncommutative algebra. As in the
quantization of classical phase-space, coordinates are replaced by
generators of the algebra~\cite{Dir26}. Since these do not commute they
cannot be simultaneously diagonalized and the space disappears. One can
argue~\cite{Mad92} that, just as Bohr cells replace
classical-phase-space points, the appropriate intuitive notion to
replace a `point' is a Planck cell of dimension given by the Planck
area. If a coherent description could be found for the structure of
space-time which were pointless on small length scales, then the
ultraviolet divergences of quantum field theory could be
eliminated. In fact the elimination of these divergences is equivalent
to coarse-graining the structure of space-time over small length
scales; if an ultraviolet cut-off $\Lambda$ is used then the theory
does not see length scales smaller than $\Lambda^{-1}$.  When a
physicist calculates a Feynman diagram he is forced to place a cut-off
$\Lambda$ on the momentum variables in the integrands.  This means
that he renounces any interest in regions of space-time of volume less
than $\Lambda^{-4}$.  As $\Lambda$ becomes larger and larger the
forbidden region becomes smaller and smaller but it can never be made
to vanish.  There is a fundamental length scale, much larger than the
Planck length, below which the notion of a point is of no practical
importance.  The simplest and most elegant, if certainly not the only,
way of introducing such a scale in a Lorentz-invariant way is through
the introduction of noncommuting space-time `coordinates'.

As a simple illustration of how a `space' can be `discrete' in some
sense and still covariant under the action of a continuous symmetry
group one can consider the ordinary round 2-sphere, which has acting on
it the rotational group $SO_3$. As a simple example of a lattice
structure one can consider two points on the sphere, for example the
north and south poles.  One immediately notices of course that by
choosing the two points one has broken the rotational invariance. It can
be restored at the expense of commutativity. The set of functions on the
two points can be identified with the algebra of diagonal $2 \times 2$
matrices, each of the two entries on the diagonal corresponding to a
possible value of a function at one of the two points. Now an action of
a group on the lattice is equivalent to an action of the group on the
matrices and there can obviously be no non-trivial action of the group
$SO_3$ on the algebra of diagonal $2 \times 2$ matrices. However if one
extends the algebra to the noncommutative algebra of all $2 \times 2$
matrices one recovers the invariance. The two points, so to speak, have
been smeared out over the surface of a sphere; they are replaced by two
cells.  An `observable' is an hermitian $2 \times 2$ matrix and has
therefore two real eigenvalues, which are its values on the two cells.
Although what we have just done has nothing to do with Planck's constant
it is similar to the procedure of replacing a classical spin which can
take two values by a quantum spin of total spin 1/2. Only the latter is
invariant under the rotation group.  By replacing the spin 1/2 by
arbitrary spin $s$ one can describe a `lattice structure' of $n = 2s+1$
points in an $SO_3$-invariant manner. The algebra becomes then the
algebra $M_n$ of $n \times n$ complex matrices and there are $n$ cells
of area $2\pi \kbar$ with
$$
n \simeq {\mbox{Vol}(S^2) \over 2\pi \kbar}.
$$ 

In general, a static, closed surface in a fuzzy space-time as we define
it can only have a finite number of modes and will be described by some
finite-dimensional algebra~\cite{GroMad92, GroPre95, GroKliPre96,
GroKliPre97a, GroKliPre97b}. Graded extensions of some of these algebras
have also been constructed~\cite{GroKliPre97c, GroRei98}. Although we
are interested in a matrix version of surfaces primarily as a model of
an eventual noncommutative theory of gravity they have a certain
interest in other, closely related, domain of physics. We have seen, for
example, that without the differential calculus the fuzzy sphere is
basically just an approximation to a classical spin $r$ by a quantum
spin $r$ with $\hbar$ in lieu of $\kbar$.  It has been extended in
various directions under various names and for various
reasons~\cite{Ber75, deWHopNic88, Hop89, BorHopSchSch91}. In order to
explain the finite entropy of a black hole it has been conjectured, for
example by 't~Hooft~\cite{tHo96}, that the horizon has a structure of a
fuzzy 2-sphere since the latter has a finite number of `points' and yet
has an $SO_3$-invariant geometry.  The horizon of a black hole might be
a unique situation in which one can actually `see' the cellular
structure of space.

It is to be stressed that we shall here modify the structure of
Minkowski space-time but maintain covariance under the action of the
Poincar\'e group. A fuzzy space-time looks then like a solid which has a
homogeneous distribution of dislocations but no disclinations. We can
pursue this solid-state analogy and think of the ordinary Minkowski
coordinates as macroscopic order parameters obtained by coarse-graining
over scales less than the fundamental scale.  They break down and must
be replaced by elements of some noncommutative algebra when one
considers phenomena on these scales.  It might be argued that since we
have made space-time `noncommutative' we ought to do the same with the
Poincar\'e group. This logic leads naturally to the notion of a
$q$-deformed Poincar\'e (or Lorentz) group which act on a very
particular noncommutative version of Minkowski space called
$q$-Minkowski space~\cite{LukNowRue92, LukRueRuh93, Cas95, AzcKulRod97,
CerWes98}. The idea of a $q$-defor\-mation goes back to
Sylvester~\cite{Syl84}. It was taken up later by Weyl~\cite{Wey50} and
Schwinger~\cite{Sch60} to produce a finite version of quantum
mechanics.

It has also been argued, for conceptual as well as practical, numerical
reasons, that a lattice version of space-time or of space is quite
satisfactory if one uses a random lattice structure or graph. The most
widely used and successful modification of space-time is in fact what is
called the lattice approximation.  From this point of view the Lorentz
group is a classical invariance group and is not valid at the
microscopic level. Historically the first attempt to make a finite
approximation to a curved manifold was due to Regge and this developed
into what is now known as the Regge calculus. The idea
is based on the fact that the Euler number of a surface can be expressed
as an integral of the gaussian curvature. If one applies this to a flat
cone with a smooth vertex then one finds a relation between the defect
angle and the mean curvature of the vertex. The latter is encoded in the
former. In recent years there has been a burst of activity in this
direction, inspired by numerical and theoretical calculations of
critical exponents of phase transitions on random surfaces. One chooses
a random triangulation of a surface with triangles of constant fixed
length, the lattice parameter. If a given point is the vertex of exactly
six triangles then the curvature at the point is flat; if there are less
than six the curvature is positive; it there are more than six the
curvature is negative. Non-integer values of curvature appear through
statistical fluctuation.  Attempts have been made to generalize this
idea to three dimensions using tetrahedra instead of triangles and
indeed also to four dimensions, with euclidean signature. The main
problem, apart from considerations of the physical relevance of a theory
of euclidean gravity, is that of a proper identification of the
curvature invariants as a combination of defect angles.  On the other
hand some authors have investigated random lattices from the point of
view of noncommutative geometry.  For an introduction to the lattice
theory of gravity from these two different points of view we refer to
the books by Ambj\o rn \& Jonsson~\cite{AmbJon97} and by
Landi~\cite{Lan97}. Compare also the loop-space approach to quantum
gravity~\cite{BaeMun94, GamPulAsh96, AshCorZap98}.

One typically replaces the four Minkowski coordinates $x^\mu$ by four
generators $q^\mu$ of a noncommutative algebra which satisfy commutation
relations of the form
\be 
[q^\mu, q^\nu] = i \kbar q^{\mu\nu}.                         \label{1.1} 
\ee 
The parameter $\kbar$ is a fundamental area scale which we shall suppose
to be of the order of the Planck area:
$$
\kbar \simeq \mu_P^{-2} = G\hbar.
$$ 
There is however no need for this assumption; the experimental bounds
would be much larger.  Equation~(\ref{1.1}) contains little
information about the algebra.  If the right-hand side does not vanish
it states that at least some of the $q^\mu$ do not commute. It states
also that it is possible to identify the original coordinates with the
generators $q^\mu$ in the limit $\kbar \to 0$:
\be
\lim_{\kbar \rightarrow 0} q^\mu = x^\mu.                     \label{1.2} 
\ee
For mathematical simplicity we shall suppose this to be the case
although one could include a singular `renormalization constant' $Z$ and
replace (\ref{1.2}) by an equation of the form
\be
\lim_{\kbar \rightarrow 0} q^\mu = Z \, x^\mu.                \label{1.3} 
\ee
If, as we shall argue, gravity acts as a universal regulator for
ultraviolet divergences then one could reasonably expect the limit
$\kbar \rightarrow 0$ to be a singular limit.

Let $\c{A}_\kbar$ be the algebra
generated in some sense by the elements $q^\mu$. We shall be here
working on a formal level so that one can think of $\c{A}_\kbar$ as
an algebra of polynomials in the $q^\mu$ although we shall implicitly
suppose that there are enough elements to generate smooth functions on
space-time in the commutative limit.  Since we have identified the
generators as hermitian operators on some Hilbert space we can identify
$\c{A}_\kbar$ as a subalgebra of the algebra of all operators on the
Hilbert space. We have added the subscript $\kbar$ to underline the
dependence on this parameter but of course the commutation relations
(1.1) do not determine the structure of $\c{A}_\kbar$, We in fact
conjecture that every possible gravitational field can be considered as
the commutative limit of a noncommutative equivalent and that the latter
is strongly restricted if not determined by the structure of the algebra
$\c{A}_\kbar$. We must have then a large number of algebras 
$\c{A}_\kbar$ for each value of $\kbar$.

Interest in Snyder's idea was revived much later when mathematicians,
notably Connes~\cite{Con86} and Woronowicz~\cite{Wor87a, Wor87b},
succeeded in generalizing the notion of differential structure to
noncommutative geometry.  Just as it is possible to give many
differential structures to a given topological space it is possible to
define many differential calculi over a given algebra. We shall use the
term `noncommutative geometry' to mean `noncommutative differential
geometry' in the sense of Connes.  Along with the introduction of a
generalized integral~\cite{ConRie87} this permits one in principle to
define the action of a Yang-Mills field on a large class of
noncommutative geometries.

One of the more obvious applications was to the study of a modified form
of Kaluza-Klein theory in which the hidden dimensions were replaced by
noncommutative structures~\cite{Mad89a, Mad89b, DubKerMad89}. In simple
models gravity could also be defined~\cite{Mad89b, Mad90} although it
was not until much later~\cite{Mou95, DubMadMasMou96, KasMadTes97a} that
the technical problems involved in the definition of this field were to
be to a certain extent overcome. Soon even a formulation of the standard
model of the electroweak forces could be given \cite{ConLot90}.  A
simultaneous development was a revival \cite{Maj88, Coq89, Mad89a} of
the idea of Snyder that geometry at the Planck scale would not
necessarily be described by a differential manifold.

One of the advantages of noncommutative geometry is that smooth,
finite examples~\cite{Mad92} can be constructed which are invariant
under the action of a continuous symmetry group. Such models
necessarily have a minimal length associated to them and quantum field
theory on them is necessarily finite~\cite{GroMad92, GroPre95,
GroKliPre96, CarWat97}.  In general this minimal length is usually
considered to be in some way or another associated with the
gravitational field. The possibility which we shall consider here is
that the mechanism by which this works is through the introduction of
noncommuting `coordinates'.  This idea has been developed by several
authors~\cite{HelTan54, Mad92, DopFreRob95, KemManMan95, FicLorWes96,
KemMan97, CerHinMadWes99a} from several points of view since the
original work of Snyder. It is the left-hand arrow of the diagram
\be
\begin{array}{ccc}
{\cal A}_\kbar &\Longleftarrow &\Omega^*({\cal A}_\kbar)\\[4pt]
\Downarrow && \Uparrow \\[4pt]
\mbox{Cut-off} &&\mbox{Gravity}
\end{array}                                                    \label{1.9}
\ee 
The $\c{A}_\kbar$ is a noncommutative algebra and the index $\kbar$
indicates the area scale below which the noncommutativity is relevant;
this would normally be taken to be the Planck area.  

The top arrow is a mathematical triviality; the 
$\Omega^*({\cal A}_\kbar)$ is a second algebra which contains
$\c{A}_\kbar$ and is what gives a differential structure to it just as
the algebra of de~Rham differential forms gives a differential structure
to a smooth manifold. There is an associated differential $d$, which
satisfies the relation $d^2 = 0$. The couple $(\Omega^*(\c{A}), d)$ is
known as a differential calculus over the algebra $\c{A}$.  The algebra
$\c{A}$ is what in ordinary geometry would determine the set of points
one is considering, with possibly an additional topological or measure
theoretic structure.  The differential calculus is what gives an
additional differential structure or a notion of smoothness. On a
commutative algebra of functions on a lattice, for example, it would
determine the number of nearest neighbours and therefore the dimension.
The idea of extending the notion of a differential to noncommutative
algebras is due to Connes~\cite{Con86, Con94, ConLot90, ConLot92} who
proposed a definition based on a formal analogy with an identity in
ordinary geometry involving the Dirac operator $\Dirac$. Let $\psi$ be a
Dirac spinor and $f$ a smooth function. Then one can write
$$
i \gamma^\alpha e_\alpha f \psi = \Dirac (f\psi) - f \Dirac \psi.
$$
Here $e_\alpha$ is the Pfaffian derivative with respect to an
orthonormal moving frame $\theta^\alpha$. This equation can be written
$$
\gamma^\alpha e_\alpha f = - i[\Dirac,  f]
$$
and it is clear that if one makes the replacement 
$$
\gamma^\alpha \mapsto \theta^\alpha
$$
then on the right-hand side one has the de~Rham differential. Inspired
by this fact, one defines a differential in the noncommutative case by
the formula
$$
df = i[F,f]
$$
where now $f$ belongs to a noncommutative algebra $\c{A}$ with a
representation on a Hilbert space $\c{H}$ and $F$ is an operator on
$\c{H}$ with spectral properties which make it look like a Dirac operator.
The triple $(\c{A}, F, \c{H})$ is called a spectral triple. It is
inspired by the $K$-cycle introduced by Atiyah~\cite{Ati69} to define a
dual to $K$-theory~\cite{Ati67}. The simplest example is obtained by
choosing $\c{A} = \b{C} \oplus \b{C}$ acting on $\b{C}^2$ by left
multiplication and 
$$
F = \left(
\begin{array}{cc}0 & 1 \\ 1 & 0\end{array}
\right).
$$
The 1-forms are then off-diagonal $2 \times 2$ complex matrices. The
differential is extended to them using the same formula as above but
with a bracket which is an anticommutator instead of a commutator. Since
$F^2 = 1$ it is immediate that $d^2 = 0$. The algebra $\c{A}$ of this
example can be considered as the algebra of functions on 2 points and
the differential can be identified with the finite-difference operator. 

One can argue~\cite{DimMad96, MadMou98, Mad99d}, not completely
successfully, that each gravitational field is the unique `shadow' in
the limit $\kbar \rightarrow 0$ of some differential structure over
some noncommutative algebra. This would define the right-hand arrow of
the diagram. A hand-waving argument can be given~\cite{BohRos33,
MadMou95} which allows one to think of the noncommutative structure of
space-time as being due to quantum fluctuations of the light-cone in
ordinary 4-dimensional space-time.  This relies on the existence of
quantum gravitational fluctuations. A purely classical argument based
on the formation of black-holes has been also
given~\cite{DopFreRob95}.  In both cases the classical gravitational
field is to be considered as regularizing the ultraviolet divergences
through the introduction of the noncommutative structure of
space-time. This can be strengthened as the conjecture that the
classical gravitational field and the noncommutative nature of
space-time are two aspects of the same thing. If the gravitational
field is quantized then presumably the light-cone will fluctuate and
any two points with a space-like separation would have a time-like
separation on a time scale of the order of the Planck time, in which
case the corresponding operators would no longer commute. So even in
flat space-time quantum fluctuations of the gravitational field could
be expected to introduce a non-locality in the theory. This is one
possible source of noncommutative geometry on the order of the Planck
scale.  The composition of the three arrows in (\ref{1.9}) is an
expression of an old idea, due to Pauli, that perturbative ultraviolet
divergences will somehow be regularized by the gravitational
field~\cite{Des57, IshSalStr71}. We refer to Garay~\cite{Gar95} for a
recent review.

One example from which one can seek inspiration in looking for examples
of noncommutative geometries is quantized phase space, which had been
already studied from a noncommutative point of view by
Dirac~\cite{Dir26}. The minimal length in this case is given by the
Heisenberg uncertainty relations or by modifications thereof
\cite{KemManMan95}. In fact in order to explain the supposed
Zitterbewegung of the electron Schr\"odinger~\cite{Sch30} had proposed to mix
position space with momentum space in order to obtain a set of
center-of-mass coordinates which did not commute. This idea has inspired
many of the recent attempts to introduce minimal lengths. We refer
to~\cite{FicLorWes96, KemMan97} for examples which are in one way or
another connected to noncommutative geometry. Another concept from
quantum mechanics which is useful in concrete applications is that of a
coherent state. This was first used in a finite noncommutative geometry
by Grosse \& Pre\v snajder~\cite{GroPre93} and later
applied~\cite{KemMan97, ChaDemPre98, ChoHinMadSte99} to the calculation
of propagators on infinite noncommutative geometries, which now become
regular 2-point functions and yield finite vacuum fluctuations.
Although efforts have been made in this direction \cite{ChoHinMadSte99}
these fluctuations have not been satisfactorily included as a source of
the gravitational field, even in some `quasi-commuta\-tive'
approximation. If this were done then the missing arrow in~(\ref{1.9})
could be drawn. The difficulty is partly due to the lack of tractable
noncommutative versions of curved spaces.

The fundamental open problem of the noncommutative theory of gravity
concerns of course the relation it might have to a future quantum theory
of gravity either directly or via the theory of `strings' and
`membranes'.  But there are more immediate technical problems which have
not received a satisfactory answer. We shall mention the problem of the
definition of the curvature. It is not certain that the ordinary
definition of curvature taken directly from differential geometry is the
quantity which is most useful in the noncommutative theory.  Cyclic
homology groups have been proposed by Connes as the appropriate
generalization to noncommutative geometry of topological invariants; the
definition of other, non-topological, invariants in not clear. It is not
in fact even obvious that one should attempt to define curvature
invariants.

There is an interesting theory of gravity, due to Sakharov and
popularized by Wheeler, called induced gravity, in which the
gravitational field is a phenomenological coarse-graining of more
fundamental fields. Flat Minkowski space-time is to be considered as a
sort of perfect crystal and curvature as a manifestation of elastic
tension, or possibly of defects, in this structure.  A deformation in
the crystal produces a variation in the vacuum energy which we perceive
as gravitational energy. `Gravitation is to particle physics as
elasticity is to chemical physics: merely a statistical measure of
residual energies.'  The description of the gravitational field which we
are attempting to formulate using noncommutative geometry is not far
from this. We have noticed that the use of noncommuting coordinates is a
convenient way of making a discrete structure like a lattice invariant
under the action of a continuous group. In this sense what we would like
to propose is a Lorentz-invariant version of Sakharov's crystal. Each
coordinate can be separately measured and found to have a distribution
of eigenvalues similar to the distribution of atoms in a crystal. The
gravitational field is to be considered as a measure of the variation of
this distribution just as elastic energy is a measure of the variation
in the density of atoms in a crystal.

We shall here accept a noncommutative structure of space-time as a
mathemetical possibility. One can however attempt to associate the
structure with other phenomena. A first step in this direction was
undoubtedly taken by Bohr \& Rosenfeld~\cite{BohRos33} when they
deduced an intrinsic uncertainty in the position of an event in
space-time from the quantum-mechanical measurement process. This idea
has been since pursued by other authors~\cite{Ahl94} and even related
to the formation of black holes~\cite{DopFreRob95, LiYon98} and to the
influence of quantum fluctuations in the gravitational
field~\cite{MadMou95, AshCorZap98}. An uncertainty relation in the
measurements of an event is one of the most essential aspects of a
noncommutative structure. The possible influence of quantum-mechanical
fluctuations on differential forms was realized some time ago by
Segal~\cite{Seg68}.  A related idea is what one might refer to as
`spontaneous lattization'. A quantum operator is a very singular
object in general and the correct definition of the space-time
coordinates, considered as quantum operators, could give rise to a
preferred set of events in space-time which has some of the aspects of
a `lattice' in the sense that each operator, has a discrete
spectrum~\cite{SchWes92, KemManMan95, FicLorWes96, KemMan97,
Kem98}. The work of Yukawa~\cite{Yuk49} and Takano~\cite{Tak99} could
be considered as somewhat similar to this, except that the fuzzy
nature of space-time is emphasized and related to the presence of
particles.  Finkelstein~\cite{Fin96} has attempted a very
philosophical derivation of the structure of space-time from the
notion of `simplicity' (in the group-theoretic sense of the word)
which has led him to the possibility of the `superposition of points',
simething very similar to noncommutativity.  We shall mention below
the attempts to derive a noncommutative structure of space-time from
string theory.

When referring to the version of space-time which we describe here we
use the adjective `fuz\-zy' to underline the fact that points are
ill-defin\-ed.  Since the algebraic structure is described by
commutation relations the qualifier `quantum' has also been
used~\cite{Sny47a, DopFreRob95, MadMou98}. This latter expression is
unfortunate since the structure has no immediate relation to quantum
mechanics and also it leads to confusion with `spaces' on which `quantum
groups' act.  To add to the confusion the word `quantum' has also been
used~\cite{GreYau97} to designate equivalence classes of ordinary
differential geometries which yield isomorphic string theories and the
word `lattice' has been used~\cite{Sny47a, FicLorWes96, tHo96} to
designate what we here qualify as `fuzzy'.

\sect{A simple example}

The algebra $\c{P}(u,v)$ of polynomials in $u = e^{ix}$, $v = e^{iy}$ is
dense in any algebra of functions on the torus, defined by the relations
$0 \leq x \leq 2\pi$, $0 \leq y \leq 2\pi$, where $x$ and $y$ are the
ordinary cartesian coordinates of $\b{R}^2$.  If one considers a square
lattice of $n^2$ points then $u^n = 1$ and $v^n = 1$ and the algebra is
reduced to a subalgebra $\c{P}_n$ of dimension $n^2$. Introduce a basis
$\ket{j}_1$, $0 \leq j \leq n-1$, of $\b{C}^n$ with 
$\ket{n}_1 \equiv \ket{0}_1$ and replace $u$ and $v$ by the operators
$$
u \ket{j}_1 = q^j \ket{j}_1, \quad v \ket{j}_1 = \ket{j+1}_1, 
\qquad q^n = 1.
$$
Then the new elements $u$ and $v$ satisfy the relations
$$
uv = q vu, \quad u^n = 1, \qquad v^n = 1
$$
and the algebra they generate is the matrix algebra $M_n$ instead of the
commutative algebra $\c{P}_n$. There is also a basis $\ket{j}_2$ in which
$v$ is diagonal and a `Fourier' transformation between the
two~\cite{Sch60}.

Introduce the forms~\cite{MadSae98}
\bea
&&\theta^1 
= - i \Big(1 - {n \over n-1} \ket{0}_2 \bra{0}\Big) u^{-1} du, 
\nonumber\\[6pt]
&&\theta^2 
= - i \Big(1 - {n \over n-1} \ket{n-1}_1 \bra{n-1}\Big) v^{-1} dv.
\nonumber
\eea
In this simple example the differential calculus can be defined by the
relations 
$$
\theta^a f = f \theta^a, \qquad
\theta^a \theta^b = - \theta^b \theta^a
$$
of ordinary differential geometry. It follows that
$$
\Omega^1(M_n) \simeq \bigoplus_1^2 M_n, \qquad d\theta^a = 0.
$$ 
The differential calculus has the form one might expect of a
noncommutative version of the torus. Notice that the differentials
$du$ and $dv$ do not commute with the elements of the algebra.

One can choose for $q$ the value 
$$
q = e^{2\pi i l/n}
$$
for some integer $l$ relatively prime with respect to $n$. The limit
of the sequence of algebras as $l/n \to \alpha$ irrational is known as
the rotation algebra or the noncommutative torus~\cite{Rie80}. This
algebra has a very rich representation theory and it has played an
important role as an example in the developement of noncommutative
geometry~\cite{ConRie87}.

\sect{Noncommutative electromagnetic theory}

The group of unitary elements of the algebra of functions on a manifold
is the local gauge group of electromagnetism and the covariant
derivative associated to the electromagnetic potential can be
expressed as a map 
\be
\c{H} \buildrel D \over \longrightarrow 
\Omega^1(V) \otimes_\c{A} \c{H}                            \label{map}
\ee
from a $\c{C}(V)$-module $\c{H}$ to the tensor product 
$\Omega^1(V) \otimes_{\c{C}(V)} \c{H}$, which satisfies a Leibniz rule
$$
D (f \psi) =  df \otimes \psi + f D\psi, \qquad 
f \in \c{C}(V), \quad \psi \in \c{H}.
$$
We shall often omit the tensor-product symbol in the following.  As far
as the electromagnetic potential is concerned we can identify $\c{H}$
with $\c{C}(V)$ itself; electromagnetism couples equally, for example,
to all four components of a Dirac spinor. The covariant derivative is
defined therefore by the Leibniz rule and the definition
$$
D\,1 = A \otimes 1 = A.
$$
That is, one can rewrite (\ref{map}) as
$$
D\psi = (\partial_\mu + A_\mu)dx^\mu \,\psi.
$$
One can study electromagnetism on a large class of noncommutative
geometries~\cite{Mad89b, DubKerMad89, ConLot90, CoqEspVai91} and there
exist many recent reviews~\cite{VarGra93, Mad99d, Kas99}. Because of the
noncommutativity however the result often looks more like nonabelian
Yang-Mills theory.

\sect{Metrics}

We shall define a metric as a bilinear map 
\be
\Omega^1(\c{A}) \otimes_{\c{A}} \Omega^1(\c{A})
\buildrel g \over \longrightarrow \c{A}.                       \label{metric}
\ee
This is a `conservative' definition, a straightforward generalization of
one of the possible definitions of a metric in ordinary differential
geometry: 
$$
g(dx^\mu \otimes dx^\nu) = g^{\mu\nu}.
$$
The usual definition of a metric in the commutative case 
is a bilinear map
$$
\c{X} \otimes_{\c{C}(V)} \c{X}
\buildrel g \over \longrightarrow \c{C}(V)
$$
where $\c{X}$ is the $\c{C}(V)$-bimodule of vector fields on $V$: 
$$
g(\partial_\mu \otimes \partial_\nu) = g_{\mu\nu}.
$$
This definition is not suitable in the noncommutative case since the
set of derivations of the algebra, which is the generalization of
$\c{X}$, has no natural structure as an $\c{A}$-module.  The linearity
condition is equivalent to a locality condition for the metric; the
length of a vector at a given point depends only on the value of the
metric and the vector field at that point. In the noncommutative case
bilinearity is the natural (and only possible) expression of locality.
It would exclude, for example, a metric in ordinary geometry defined
by a map of the form
$$
g(\alpha, \beta)(x) = \int_V g_x(\alpha_x, \beta_y) G(x,y) dy.
$$
Here $\alpha, \beta \in \Omega^1(V)$ and $g_x$ is a metric on the
tangent space at the point $x \in V$. The function $G(x,y)$ is an
arbitrary smooth function of $x$ and $y$ and $dy$ is the measure on $V$
induced by the metric.

Introduce a bilinear flip $\sigma$: 
\be
\Omega^1(\c{A}) \otimes_{\c{A}} \Omega^1(\c{A})
\buildrel \sigma \over \longrightarrow
\Omega^1(\c{A}) \otimes_{\c{A}} \Omega^1(\c{A})              \label{flip}
\ee
We shall say that the metric is symmetric if
$$
g \circ \sigma \propto g.
$$
Many of the finite examples have unique metrics~\cite{MadMasMou95} as do
some of the infinite ones~\cite{CerHinMadWes99a}. Other definitions of a
metric have been given, some of which are similar to that given above
but which weaken the locality condition~\cite{CerHinMadWes99b} and
one~\cite{ConLot92} which defines a metric on the associated space of
states.

\sect{Linear Connections}

An important geometric problem is that of comparing vectors and forms
defined at two different points of a manifold. The solution to this
problem leads to the concepts of a connection and covariant derivative.
We define a linear connection as a covariant derivative
$$
\Omega^1(\c{A}) \buildrel D \over \longrightarrow 
\Omega^1(\c{A}) \otimes_\c{A} \Omega^1(\c{A})
$$
on the $\c{A}$-bimodule $\Omega^1(\c{A})$ with an extra right Leibniz
rule
$$
D(\xi f) = \sigma (\xi \otimes df) + (D\xi) f
$$
defined using the flip $\sigma$ introduced in (\ref{flip}). In ordinary
geometry the map 
$$
D(dx^\lambda) = - \Gamma^\lambda_{\mu\nu} dx^\mu \otimes dx^\nu
$$
defines the Christophel symbols.

We define the torsion map
$$
\Theta: \Omega^1(\c{A}) \rightarrow \Omega^2(\c{A})
$$
by $\Theta = d - \pi \circ D$. It is left-linear. A short calculation yields 
$$
\Theta(\xi)f - \Theta(\xi f) = \pi \circ (1 + \sigma) (\xi \otimes df).
$$
We shall impose the condition
\be
\pi \circ (\sigma + 1) = 0
\ee
on $\sigma$. It could also be considered as a condition on the product
$\pi$. In fact in ordinary geometry it is the definition of $\pi$; a
2-form can be considered as an antisymmetric tensor. Because of this
condition the torsion is a bilinear map.  Using $\sigma$ a reality
condition on the metric and the linear connection can be
introduced~\cite{FioMad98}. In the commutative limit, when it exists,
the commutator defines a Poisson structure, which normally would be
expected to have an intimate relation with the linear connection. This
relation has only been studied in very particular situations~\cite{Mad99a}.

\sect{Gravity}

The classical gravitational field is normally supposed to be described
by a torsion-free, metric-compatible linear connection on a smooth
manifold. One might suppose that it is possible to formulate a
noncommutative theory of (classical/quantum) gravity by replacing the
algebra of functions by a more general algebra and by choosing an
appropriate differential calculus. It seems however difficult to
introduce a satisfactory definition of local curvature and the
corresponding curvature invariants~\cite{CunQui95, DubMadMasMou95,
DabHajLanSin96}. One way of circumventing this problem is to consider
classical gravity as an effective theory and the Einstein-Hilbert action
as an induced action.  We recall that the classical gravitational action
is given by
$$
S[g] = \mu_P^4 \Lambda_c + \mu_P^2 \int R.
$$
In the noncommutative case there is a natural definition of the
integral~\cite{ConRie87, Con88, Con94} but there does not seem to be a
natural generalization of the Ricci scalar. One of the problems is the 
fact that the natural generalization of the curvature form is in
general not right-linear in the noncommutative case. The Ricci scalar
then will not be local. One way of circumventing these problems is to
return to an old version of classical gravity known as induced
gravity~\cite{Sak67, Sak75}. The idea is to identify the gravitational
action with the quantum corrections to a classical field in a curved
background. If $\Delta[g]$ is the operator which describes the
propagation of a given mode in presence of a metric $g$ then one finds
that, with a cut-off $\Lambda$, the effective action is given by
$$
\Gamma[g] \propto \tr \log \Delta[g] \simeq
\Lambda^4 \hbox{Vol}(V)[g] + \Lambda^2 S_1[g] +
(\log \Lambda) S_2[g] + \cdots.
$$
If one identifies $\Lambda = \mu_P$ then one finds that $S_1[g]$ is the
Einstein-Hilbert action. A problem with this is that it can be only
properly defined on a compact manifold with a metric of euclidean
signature and Wick rotation on a curved space-time is a rather delicate
if not dubious procedure. Another problem with this theory, as indeed
with the gravitational field in general, is that it predicts an extremely
large cosmological constant.  The expression $\tr \log \Delta[g]$ has a
natural generalization to the noncommutative case~\cite{KalWal95,
AckTol96, ChaCon96}.

We have defined gravity using a linear connection, which required the
full bimodule structure of the $\c{A}$-module of 1-forms. One can argue
that this was necessary to obtain a satisfactory definition of locality
as well as a reality condition. It is possible to relax these
requirements and define gravity as a Yang-Mills field~\cite{ChaFelFro93,
LanNguWal94, ChaFroGra95, FroGraRec97} or as a couple of left and right
connections~\cite{CunQui95, DabHajLanSin96}. If the algebra is
commutative (but not an algebra of smooth functions) then to a certain
extent all definitions coincide~\cite{Lan97, BalBimLanLizTeo98}.

\sect{Regularization}

Using the diagram~(\ref{1.9}) we have argued that gravity regularizes
propagators in quantum field theory through the formation of a
noncommutative structure. Several explicit examples of this have been
given in the literature~\cite{Sny47b, Mad92, DopFreRob95, KemManMan95,
KraWul99, ChoHinMadSte99}. In particular an energy-momentum tensor
constructed from regularized propagators ~\cite{ChoHinMadSte99} has
been used as a source of a cosmological solution. The propagators
appear as if they were derived from non-local theories on ordinary
space-time~\cite{Yuk49, PaiUhl50, KatYuk68, Tak99}. We required that
the metric that we use be local in the sense that the map
(\ref{metric}) is bilinear with respect to the algebra. One could say
that the theory is as local as the algebra will permit. However, since
the algebra is not an algebra of points this means that the theory
appears to be non-local as an effective theory on a space-time
manifold.

\sect{Kaluza-Klein theory}

We mentioned in the Introduction that one of the first, obvious
applications of noncommutative geometry is as an alternative hidden
structure of Kaluza-Klein theory. This means that one leaves space-time
as it is and one modifies only the extra dimensions; one replaces their
algebra of functions by a noncommutative algebra, usually of finite
dimension to avoid the infinite tower of massive states of traditional
Kaluza-Klein theory. Because of this restriction and because the extra
dimensions are purely algebraic in nature the length scale associated
with them can be arbitrary~\cite{MadMou93}, indeed as large as the
Compton wave length of a typical massive particle. 

The algebra of Kaluza-Klein theory is therefore, for example, a product
algebra of the form
$$
\c{A} = \c{C}(V) \otimes M_n.
$$
Normally $V$ would be chosen to be a manifold of dimension four, but
since much of the formalism is identical to that of the
$M$(atrix)-theory of $D$-branes~\cite{BanFisSheSus97, GanRamTay97,
ConDouSch98}.  For the simple models with a matrix extension one can use
as gravitational action the Ein\-stein-Hilbert action in `dimension'
$4+d$, including possibly Gauss-Bonnet terms~\cite{Mad90, MadMou93,
Mad99d, MadMou95, KehMadMouZou95}. For a more detailed review we refer
to a lecture~\cite{MadMou96} at the 5th Hellenic school in Corfu.

\sect{Quantum groups and spaces}

The set of smooth functions on a manifold is an algebra. This means
that from any function of two variables one can construct a function
of one by multiplication. If the manifold happens to be a Lie group
then there is another operation which to any function of one variable
constructs a function of two. This is called co-multiplication and is
usually written $\Delta$: 
$$
(\Delta f)(g_1, g_2) = f(g_1g_2).
$$
It satisfies a set of consistency conditions with the product.  Since
the expression `noncommutative group' designates something else the
noncommutative version of an algebra of smooth functions on a Lie
group has been called a `quantum group'. It is neither `quantum' nor
`group'. The first example was found by Kulish \&
Reshetikhin~\cite{KulRes83} and by Sklyanin~\cite{Sky82}. A systematic
description was first made by Woronowicz~\cite{Wor80}, by
Jimbo~\cite{Jim85}, Manin~\cite{Man88, Man89} and
Drinfeld~\cite{Dri86}. The Lie group $SO(n)$ acts on the space
$\b{R}^n$; the Lie group $SU(n)$ acts on $\b{C}^n$. The `quantum'
versions $SO_q(n)$ and $SU_q(n)$ of these groups act on the `quantum
spaces' $\b{R}^n_q$ and $\b{C}^n_q$. These latter are noncommutative
algebras with special covariance properties. The first differential
calculus on a quantum space was constructed by Wess \&
Zumino~\cite{WesZum90}.  There is an immense literature on quantum
groups and spaces, from the algebraic as will as geometric point of
view. We have included some of it in the bibliography. We mention in
particular the collection of articles edited by Doebner \&
Hennig~\cite{DoeHen90} and Kulish~\cite{Kul91} and the introductory text
by Kassel~\cite{Kass95}.

\sect{Mathematics}

At a more sophisticated level one would have to add a topology to the
algebra. Since we have identified the generators as hermitian
operators on a Hilbert space, the most obvious structure would be that
of a von~Neumann algebra. We refer to Connes~\cite{Con94} for a
description of these algebras within the context of noncommutative
geometry. A large part of the interest of mathematicians in
noncommutative geometry has been concerned with the generalization of
topological invariants~\cite{Con86, CunQui95, Mos97} to the
noncommutative case. It was indeed this which lead Connes to develop
cyclic cohomology. Connes~\cite{ConRie87, Con88} has also developed
and extended the notion of a Dixmier trace on certain types of
algebras as a possible generalization of the notion of an integral.
The representation theory of quantum groups is an active field of
current interest since the pioneering work of Woronowicz~\cite{Wor91}.
For a recent survey we refer to the book by Klymik \&
Schm\"udgen~\cite{KlySch97}. Another interesting problem is the
relation between differential calculi covariant under the (co-)action
of quantum groups and those constructed using the spectral-triple
formalism of Connes. Although it has been known for some
time~\cite{DubMadMasMou95, DimMad96, GeoMadMasMou97, ChoMadPar98,
FioMad99} that many if not all of the covariant calculi have formal
Dirac `operators' it is only recently that mathematicians have
considered~\cite{Sch99a, Sch99b} to what extent these `operators' can
be actually represented as real operators on a Hilbert space and to
what extent they satisfy the spectral-triple conditions.

\sect{String Theory}

Last, but not least, is the possible relation of noncommutative geometry
to string theory. We have mentioned that since noncommutative geometry
is pointless a field theory on it will be divergence-free. In particular
monopole configurations will have finite energy, provided of course that
the geometry in which they are constructed can be approximated by a
noncommutative geometry, since the point on which they are localized has
been replaced by an volume of fuzz, This is one characteristic that it
shares with string theory. Certain monopole solutions in string theory
have finite energy~\cite{GauGomTow99} since the point in space (a
$D$-brane) on which they are localized has been replaced by a throat to
another `adjacent' $D$-brane. 

In noncommutative geometry the string is replaced by a certain finite
number of elementary volumes of `fuzz', each of which can contain one
quantum mode. Because of the nontrivial commutation relations the `line'
$\delta q^\mu = q^{\mu\prime} - q^\mu$ joining two points
$q^{\mu\prime}$ and $q^\mu$ is quantized and can be
characterized~\cite{ChoHinMadSte99} by a certain number of creation
operators $a_j$ each of which creates a longitudinal displacement. They
would correspond to the rigid longitudinal vibrational modes of the
string. Since it requires no energy to separate two points the string
tension would be zero. This has not much in common with traditional
string theory.

We mentioned in the previous section that noncommutative Kaluza-Klein
theory has much in common with the $M$(atrix) theory of
$D$-branes. What is lacking is a satisfactory supersymmetric
extension.  Finally we mention that there have been speculations that
string theory might give rise naturally to space-time uncertainty
relations \cite{LiYon98} and that it might also give
rise~\cite{JevRam99} to a noncommutative theory of gravity. More
specifically there have been attempts~\cite{DouHul98, Dou99, Sch99} to
relate a noncommutative structure of space-time to the quantization of
the open string in the presence of a non-vanishing $B$-field.

\section*{Acknowledgments} The author would like to thank the 
Max-Planck-Institut f\"ur Physik in M\"unchen for financial support
and J. Wess for his hospitality there.

\tolerance=1000
\pretolerance=2000
\hfuzz=4pt
\hbadness=3000
\hbadness = 10000
\parskip=0pt

%
%

\newcommand\aut[2]{#1\ #2,}
\newcommand\autb[4]{#1\ #2,\ #3\ #4,}
\newcommand\autc[6]{#1\ #2,\ #3\ #4,\ #5\ #6,}
\newcommand\autd[8]{#1\ #2,\ #3\ #4,\ #5\ #6,\ #7\ #8,}

\newcommand\ed[2]{#1\ #2\ (Ed.),}
\newcommand\edb[4]{#1\ #2,\ #3\ #4\ (Eds.),}
\newcommand\edc[6]{#1\ #2,\ #3\ #4,\ #5\ #6\ (Eds.),}
\newcommand\edd[8]{#1\ #2,\ #3\ #4,\ #5\ #6,\ #7\ #8\ (Eds.),}

\newcommand\epr[4]{``#1'',\ #2,\ {\tt #3#4}}

\newcommand\jour[5]{``#1'',\ {\it #2\ }{\bf #3}\ (#4)\ #5}

\newcommand\confj[6]{``#1'',\ #2,\ {\it #3\ }{\bf #4}\ (#5)\ #6}

\newcommand\series[5]{{\it #1},\ #2\ No.\ #3,\ #4,\ #5}

\newcommand\book[3]{{\it #1},\ #2,\ #3}

\newcommand\confb[4]{{\it #1},\ #2,\ #3,\ #4}

\newcommand\adp  {Adv.\ Phys.}
\newcommand\agag {Ann. Global Anal. Geom.} 
\newcommand\ajm  {Amer. J. Math.} 
\newcommand\am   {Ann.\ of\ Math.}
\newcommand\ap   {Ann.\ Phys.}
\newcommand\arnps{Ann.\ Rev.\ Nucl.\ Part.\ Sci.}
\newcommand\atmp {Adv.\ Theor.\ Math.\ Phys.}
\newcommand\bb   {Berliner Ber.}
\newcommand\cm   {Contemp.\ Math.}
\newcommand\cmp  {Comm.\ Math.\ Phys.}
\newcommand\cpc  {Comput.\ Phys.\ Commun.}
\newcommand\cqg  {Class.\ and Quant.\ Grav.}
\newcommand\cras {C.\ R. Acad. Sci. Paris} 
\newcommand\dokl {Doklady Akad. Nauk. S.S.S.R.}
\newcommand\epj  {Euro.\ Phys.\ J.\ C}
\newcommand\grg  {Gen. Rel. Grav.}
\newcommand\faa  {Funct. Anal. Appl.}
\newcommand\fdp  {Fortschritte der Phys.}
\newcommand\hpa  {Helv.\ Phys.\ Acta}
\newcommand\ihes {Publications of the I.H.E.S.}
\newcommand\ijm  {Israel J. Math.} 
\newcommand\ijmp {Int.\ J.\ Mod.\ Phys.\ A}
\newcommand\ijmps{Int.\ J.\ Mod.\ Phys.\ A (Proc. Suppl.)}
\newcommand\ijtp {Int. J. Theor. Phys.} 
\newcommand\inm  {Invent. Math.} 
\newcommand\jmp  {J.\ Math.\ Phys.}
\newcommand\jetp {Sov.\ Phys.\ JETP}
\newcommand\jetpl{JETP Lett.}
\newcommand\ja   {Jour. of Alg.}
\newcommand\jhep {J. High Energy Phys.}
\newcommand\jgp  {J.\ Geom.\ Phys.}
\newcommand\joth {J.\ Operator Theory }
\newcommand\jpa  {J.\ Phys.\ A:\ Math.\ Gen.}
\newcommand\jsm  {J.\ Soviet\ Math.}
\newcommand\kdv  {Kgl. Danske. Vidensk. Selskab. Mat-fyz. Meddelser}
\newcommand\lmp  {Lett.\ Math.\ Phys.}
\newcommand\lnm  {Lect.\ Notes\ in\ Math.}
\newcommand\lnp  {Lect.\ Notes\ in\ Phys.}
\newcommand\mpla {Mod.\ Phys.\ Lett.\ A}
\newcommand\nc   {Nuovo Cim.}
\newcommand\np   {Nucl.\ Phys.}
\newcommand\npps {Nucl.\ Phys.\ (Proc.\ Suppl.)}
\newcommand\pb   {Physica\ B}
\newcommand\pjm  {Pacific J. Math.} 
\newcommand\pl   {Phys.\ Lett.}
\newcommand\pmb  {Progress in Math. (Birkh\"auser)}
\newcommand\pnas {Proc. Nat. Acad. Sci. (USA)}
\newcommand\pcps {Proc. Camb. Phil. Soc.}
\newcommand\ppnp {Prog.\ Part.\ Nucl.\ Phys.}
\newcommand\pr   {Phys.\ Rev.}
\newcommand\prep {Phys.\ Rep.}
\newcommand\prl  {Phys.\ Rev.\ Lett.}
\newcommand\prs  {Proc.\ Roy.\ Soc.}
\newcommand\ptp  {Prog.\ Theor.\ Phys.}
\newcommand\ptps {Prog.\ Theor.\ Phys.\ (Suppl.)}
\newcommand\rmp  {Rev.\ Mod.\ Phys.}
\newcommand\rmap {Rev.\ Math.\ Phys.}
\newcommand\rims {Publ. RIMS, Kyoto Univ.} 
\newcommand\rpmp {Rep.\ on\  Mod.\ Phys.}
\newcommand\rpmap{Rep.\ on\  Math.\ Phys.}
\newcommand\rpp  {Rep. Prog. Phys.}
\newcommand\sjnp {Sov.\ J.\ Nucl.\ Phys.}
\newcommand\smd  {Sov. Math. Dokl.}
\newcommand\smf  {Soc. Math. Fran\c{c}ais}
\newcommand\spaw {Sitzungber. Preuss. Akad. Wiss. Phys.-Math. Kl.}
\newcommand\tmf  {Teor. i Mat. Fiz.} 
\newcommand\topo {Topology}
\newcommand\umn  {Uspekhi Math. Nauk.}
\newcommand\yf   {Yad.\ Fiz.}
\newcommand\zp   {Z.\ Physik}
\newcommand\zpc  {Z.\ Physik C - Particles and Fields}
\newcommand\zetf {Zh.\ Eksp.\ Teor.\ Fiz.}

\newcommand\ibid {{\it ibid}}
\newcommand\idem {{\it idem}}

\newcommand{\alggeom}{alg-geom/}
\newcommand{\astroph}{astro-ph/}
\newcommand{\grqc}{gr-qc/}
\newcommand{\hepth}{hep-th/}
\newcommand{\ma}{math/}
\newcommand{\mathph}{math-ph/}
\newcommand{\qalg}{q-alg/}
\newcommand{\quantph}{quant-ph/}

\newcommand\acp {Academic Press}
\newcommand\ah  {Adam Hilger, Bristol}
\newcommand\ams {Amer. Math. Soc., Providence, Rhode Island}
\newcommand\av  {Akademie-Verlag, Berlin}
\newcommand\aw  {Addison-Wesley Publishing Co.}
\newcommand\bgt {B.G. Teubner, Stuttgart}
\newcommand\bir {Birkh\"auser, Basel}
\newcommand\bc  {Benjamin/Cummings}
\newcommand\bri {E.J. Brill, Leiden}
\newcommand\cam {Cambridge University Press}
\newcommand\clp {Clarendon Press, Oxford}
\newcommand\dov {Dover, New York}
\newcommand\fre {W.H. Freeman and Company}
\newcommand\her {Hermann, Paris}
\newcommand\hp  {Heron Press, Sofia}
\newcommand\ip  {International Press}
\newcommand\jws {John Wiley \& Son}
\newcommand\kap {Kluwer Academic Publisher}
\newcommand\mh  {McGraw-Hill, Princeton}
\newcommand\nh  {North-Holland Publishing, Amsterdam}
\newcommand\nsp {Nova Science Publishers, USA}
\newcommand\oup {Oxford University Press}
\newcommand\pp  {Plenum Press, New York}
\newcommand\pup {Princeton University Press}
\newcommand\rp  {D. Reidel Publishing Company}
\newcommand\sv  {Springer-Verlag, Heidelberg}
\newcommand\vv  {Friedr. Vieweg \& Sohn Verlag, Braunschweig/Wiesbaden}
\newcommand\wg  {Walter de Gruyter, Berlin}
\newcommand\ws  {World Scientific Publishing}
\newcommand\wab {W.A. Benjamin, New York}

\section*{}

What follows constitutes in no way a complete bibliography of
noncommutative geometry. It is strongly biased in favour of the
author's personal interests and the few subjects which were touched
upon in the text.

\end{document}